# Time-domain simulator of Josephson junctions based on the BCS theory

L. Iwanikow, P. Febvre, *Senior Member, IEEE*

*Abstract*—We realized a time-domain simulator based on the electrodynamics of Cooper pairs and quasiparticles in Josephson junctions. The tool, based on the charge carriers' densities of states described by the Werthamer and Harris formalisms of Bardeen-Cooper-Schrieffer (BCS) theory, allows to analyze the behavior of current- or voltage-controlled Josephson junction-based circuits for any signal waveform, at a physical temperature comprised between the absolute temperature and the critical temperature. The simulator can account for *I-V* curve hysteresis depending on the McCumber parameter, or Shapiro and photo-assisted steps in the presence of a monochromatic microwave signal. We used the simulator to assess the behavior of MgB$_2$-based Josephson junctions at THz frequencies by taking into account the presence of the two anisotropic gaps of MgB2.

*Index Terms*—BCS theory, Josephson junction, time-domain simulation, quasiparticle, MgB2

## I. Introduction

Simulators are crucial tools for the design of superconducting electronic circuits, whether they are analog, digital or quantum. The accuracy, reliability and capabilities of simulation tools are increasingly important for the development of more complex Single Flux Quantum (SFQ)-based and superconductor-based RF applications for which higher frequencies imply dissipative and dispersive mechanisms near the energy gap of superconductors.

Software like JSIM [1], PSCAN [2], WRSpice [3] or JoSIM [4] do not always take into account all electrical phenomena caused by the dynamics of Cooper pair and quasiparticles tunneling in a Josephson junction. The time-domain simulator presented here aims at giving a precise description of the physical behavior of a Josephson junction based on phase and current quantities. The voltage is obtained as the derivative of the phase. It is based on the calculation of the tunneling current through the Josephson junction barrier, based on the densities of states of Cooper pairs and quasiparticles described by the BCS theory [5] for both superconducting electrodes.

Following the prediction that a tunneling supercurrent can pass through a thin insulating barrier between two superconducting electrodes by Josephson [6-7], several models attempting to explain this phenomenon in accordance with BCS theory emerged. Werthamer formalism described the tunneling of Cooper pairs and quasiparticles for a voltage-biased junction at *T = 0 K* [8]. Later, Larkin and Ovchinnikov have taken up these results, adding the influence of the temperature in the calculations [9]. This work has been revised under a different formalism by Harris who further detailed the meaning of the different parts that make up the tunneling current [10-11].

Based on these various models, our simulator is applicable for weak-linked superconductor-insulator-superconductor (SIS) Josephson junctions, provided that the superconducting materials used in the junction are described by the BCS theory and that the junction size is small compared to the Josephson length (which is typically 4 µm for a niobium-based junction of 10 kA/cm$^2$ current density), this in order to keep the current distribution about constant over the junction area. Under these conditions, we can consider that the whole electrodynamics of the Josephson junctions is taken into consideration. To prove that this is indeed the case, we simulate the effects of an incoming RF radiation on the junction *I-V* curve and deduce the relation between the height of photo-assisted steps and the junction RF voltage induced by the radiation [12].

## II. Electrodynamics of a Josephson junction

### A. BCS description of the densities of states of charge carriers

Electrodynamics of a Josephson junction can be obtained through the calculation of frequency spectra of the tunneling currents, called kernels, for both charge carriers, i.e. Cooper pairs and quasiparticles, for superconductors described by the BCS theory these spectra are obtained by the calculation of the probability of passage of charges by tunneling effect through the Josephson junction insulating barrier. Currents are associated with the densities of energy states of charge carriers in each superconducting electrode. Frequency is the natural quantity to calculate tunneling currents because they are directly linked to the energy states of both superconductors of the junction.

### B. Frequency kernels

Our work is a direct application of the expression of kernels described by Werthamer [8], Larkin & Ovchinnikov [9] and Harris [10-11]. The calculation of kernels is done by computing the integral over all frequencies of the charge carriers tunneling through the barrier.

Calculations rely on physical and electrical quantities such as the barrier resistance, the critical current of the junction, the critical temperatures of the superconducting materials, and the

Manuscript receipt and acceptance dates will be inserted here. This work is supported by the Agence de l'Innovation de Défense (AID), the Direction Générale de l'Armement (DGA), the Centre National d'Etudes Spatiales (CNES) and the Office of the Director of National Intelligence (ODNI), the Intelligence Advanced Research Projects Activity (IARPA), via the U.S. Army Research Office Grant W911NF-17-1-0120.

L. Iwanikow and P. Febvre are with IMEP-LAHC, University Savoie Mont Blanc, Le Bourget du Lac, France (e-mail: lucas.iwanikow@univ-smb.fr; pascal.febvre@univ-smb.fr).

physical temperature of the device. The frequency kernels are complex functions since they take into account the retarded response of charge carriers to a given forced solicitation, that becomes dominant at the gap frequency which is a characteristic frequency of the Josephson junction.

Fig. 1-a shows the frequency kernels for a niobium-based Josephson junction. We see that their imaginary part is an odd function of voltage due to the junction symmetry, while their real part is even since real and imaginary parts are connected through Kramers-Kronig relations [11] for causality reasons.

### C. Analytical fits of kernels

To increase the speed and flexibility of calculations we fit the kernels using complex Lorentzian sums based on the OSZ algorithm suggested in [13] and by Jablonski [14], as illustrated by Eq. 1-a and Eq. 1-b. Such a choice allows to keep the essential information on the junction dynamics contained in the kernels, while simplifying the calculation of currents in the time domain.

$$J_p(\omega) = \sum_n \left[ \frac{A_n}{q_n + j\omega} + \frac{A_n^*}{q_n^* + j\omega} \right] \quad (1.a)$$

$$J_{qp}(\omega) = j\omega + \sum_n \left[ \frac{B_n}{r_n + j\omega} + \frac{B_n^*}{r_n^* + j\omega} \right] \quad (1.b)$$

Fits were made in Python 3 using the LMFIT library [15], which gives fitting parameters showing a good agreement with the theoretical curves. An illustration of the concordance is given in Fig. 1-b.

## III. TIME-DOMAIN SIMULATIONS

### A. Josephson and quasiparticles currents

The use of the information contained in the frequency kernels of the simulated Josephson junction has been highlighted by Gayley [16] and Harris [11]. Equations (2.a) and (2.b) exhibit the time-domain evolution of the current due to Cooper pairs and quasiparticles respectively.

$$I_p(t) = \Im\left\{ W(t) \int_{-\infty}^{t} J_p(t-\tau) U(\tau) d\tau \right\} \quad (2.a)$$

$$I_{qp}(t) = \Im\left\{ W^*(t) \int_{-\infty}^{t} J_{qp}(t-\tau) U(\tau) d\tau \right\} \quad (2.b)$$

with $W(t) = \exp(j\varphi(t)/2)$. Thus the inverse Fourier transforms of the fits of kernel currents give Eq. 3-a and Eq. 3-b for Cooper pairs and quasiparticles respectively.

$$J_p(t) = \theta(t) \sum_n [A_n \exp(-q_n t) + A_n^* \exp(-q_n^* t)] \quad (3.a)$$

$$J_{qp}(t) = \theta(t) \left\{ \delta'(t) + \sum_n [B_n \exp(-r_n t) + B_n^* \exp(-r_n^* t)] \right\} \quad (3.b)$$

with $\theta(t)$ being the Heaviside function in time-domain.

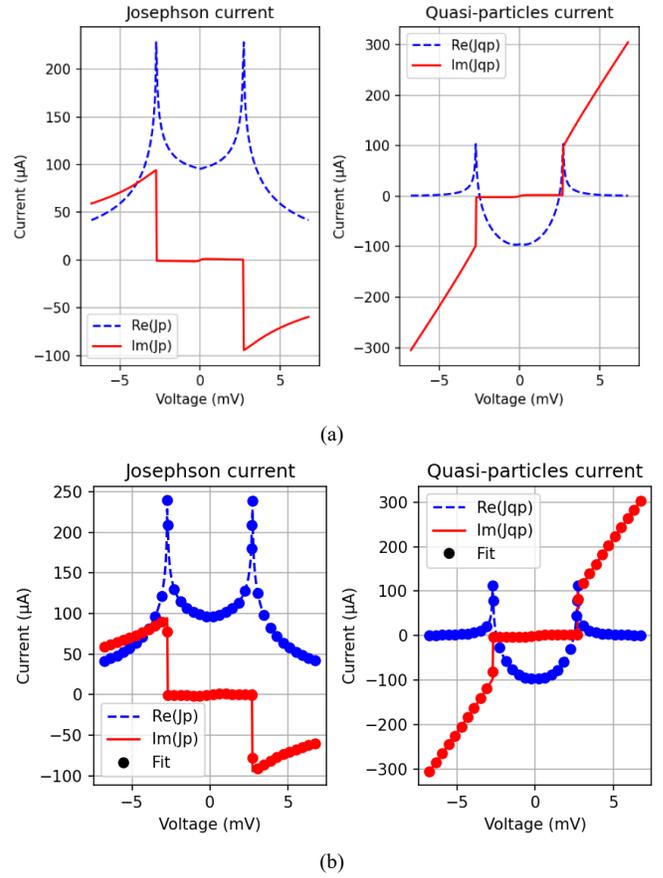

Fig. 1. (a) Frequency kernels for a niobium-based Josephson junction with a critical current $I_C$ = 100 µA calculated for $T$ = 4.2 K. The value of the critical current of the junction can be interpreted as the real part of the Cooper pairs current at zero voltage. (b) Fits of the simulated Josephson junction illustrated in Fig. 1-a, showing a good agreement with the theoretical curves.

### B. Phase, voltage, or current control

The application to the Josephson junction of a user-defined voltage or phase waveform is straightforward from equations (2.a) and (2.b). On the other hand, if one wishes to simulate a current-controlled junction and observe the evolution of the voltage across it, it is necessary to use an algorithm such as Runge-Kutta or Newton-Raphson. This second algorithm is the one that has been deployed for our simulator.

Time-domain simulations are carried out by initializing all our variables at time $t = 0$ and calculating current increments time step by time step.

### C. I-V curves

Junction I-V curves can be obtained by considering only the Josephson current, the quasiparticle current or both. To take into account some phenomena described by the standard Resistively and Capacitively Shunted Junction (RCSJ) model [17-18], the junction capacitance can be added in parallel, as well as a shunt resistor, which is of interest in the context of digital circuit simulation.

I-V curves are obtained by sweeping the junction bias current $I_b$. For each bias current a time-domain simulation calculates the voltage evolution at the junction terminals over a given time until convergence. By averaging the voltage over this time interval, one obtains a point in the *I-V* curve.

### D. Effect of an RF radiation

In order to simulate the effect of an RF radiation on the behavior of a junction, shown in Fig. 2, calculations are based on a current generator waveform given by $I(t) = I_{DC} + I_{RF} \sin(\omega t)$ where $\omega$ corresponds to the angular frequency of the RF signal.

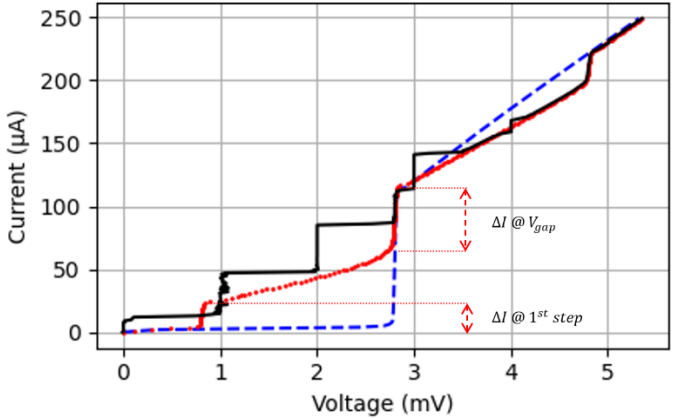

Fig. 2. Simulated *I-V* curves for quasiparticles of a niobium-based Josephson junction with $I_C = 100$ µA irradiated by a monochromatic microwave signal with *f* = 484 GHz, with (black solid line) and without (red dotted line) the Cooper pairs contribution respectively. The blue dashed line is obtained for quasiparticles only and without microwave signal. One can see the first four Shapiro steps with continuous voltage values, with a voltage spacing of 1 mV each. One can also observe the first photo-assisted step, starting from the gap at 2.8 mV, with a width of *2* mV.

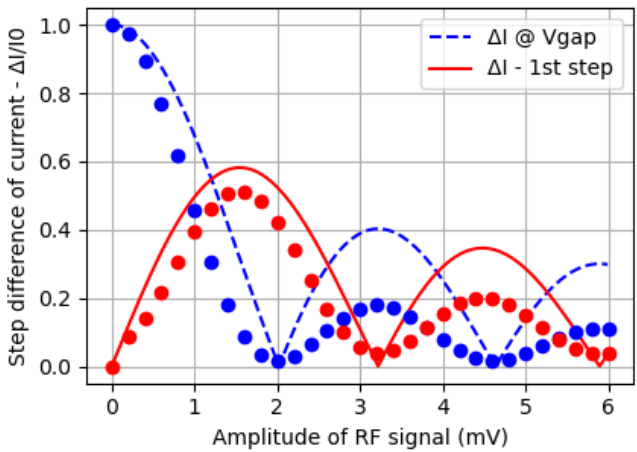

Fig. 3. Evolution of the normalized amplitude (current discontinuity divided by the current $I_0 = V_g/R_n$ of the junction) of the discontinuity at the gap voltage and the first photo-assisted step as a function of RF voltage, for a niobium-based Josephson junction with $I_C = 100$ µA irradiated at *f* = 203 GHz. Lines (dashed and solid) represent the theoretical absolute value of Bessel function of 1st kind of order 0 (blue) and 1 (red). Dots illustrate the simulated evolution of the amplitude of the gap (blue) and of the height of the first photo-assisted step (red).

Shapiro proved that the presence of a monochromatic microwave signal caused constant voltage steps to appear in the *I-V* curve of a Josephson junction [19]. Two types of steps are to be distinguished, the Shapiro steps being the response to the RF signal of Cooper pairs, while the photo-assisted steps are the response of quasiparticles. The width of these steps is closely related to the frequency of the applied radiation.

Fig. 3 shows the dependence of the height of the first photo-assisted step on the amplitude of the applied microwave voltage. As described in the literature [12], the height of the step seems to follow a Bessel function of the first kind of the order 1. The same behavior has been observed for the height of the *n-th* photo-assisted step following a Bessel function of the first kind of the order *n*.

## IV. MgB₂-BASED JOSEPHSON JUNCTION SIMULATIONS

### A. Frequency kernels

Magnesium diboride ($MgB_2$) is a material with a relatively high critical temperature (39 K) compared to niobium, that can have interesting applications especially at high frequencies, such as RSFQ logic. Although it has two anisotropic gaps [20-21], this material can be described by the BCS theory.

The description of this double gap is made with the approximation that the densities of states can be expressed as a weighted sum of two densities of states, one for each half-gap, as suggested by Giubileo [22].

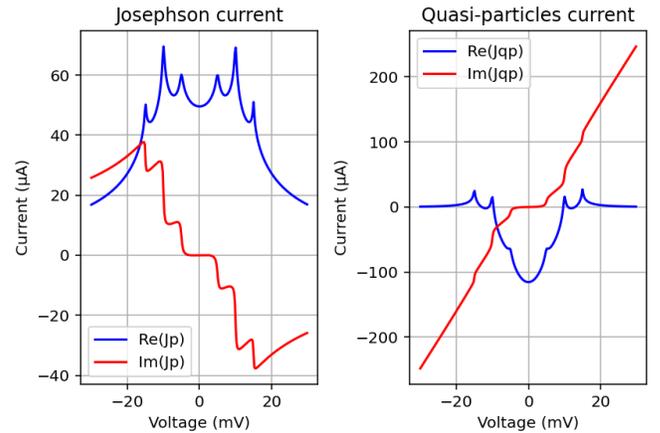

Fig. 4. Frequency kernels for a $MgB_2$-based Josephson junction with a critical current $I_C = 50$ µA calculated for *T = 4.2 K*. We can observe singularities for voltages equal to *± 5 mV, ± 10 mV* and *± 15 mV*.

Fig. 4 illustrates an example of frequency kernel for a $MgB_2$-based Josephson junction with a critical current $I_C = 50$ µA calculated for *T = 4.2 K*. The presence of three peaks is a direct consequence of the two gaps for $MgB_2$. The first peak is due to the interaction between the narrow half-gaps ($\Delta_S$ = *2.5* meV), while the third peak is the combination of the large half-gaps ($\Delta_L$ = *7.5* meV). The second peak results from the interaction of the narrow half-gap from one side of the junction and the large half-gap from the other side.

### B. Simulations of SFQ pulses generation

Magnesium diboride having a large half-gap $\Delta_L$, it can be an interesting material for high-frequency applications close to the gap frequency of $f = 3.6$ THz, like for heterodyne receivers (superconducting mixers, or Hot Electron Bolometers (HEBs)) for THz astronomy [23-26] or security applications, or Superconducting Nanowire Single-Photon Detectors (SNSPDs) [27]. Besides, its higher critical temperature allows to work around 20K which greatly relaxes cryogenic constraints, especially for space-based instruments. In an attempt to investigate the simulator for digital applications, we used our model to describe the behavior of a current-biased Josephson junction for the generation of an SFQ signal, by biasing it above its critical current.

Fig. 5-a shows the simulation of a resistively shunted MgB$_2$-based Josephson junction biased with a DC current slightly over the critical current at a clock rate of 244 GHz, considering the Cooper pairs contribution, with or without the contribution of quasiparticles. $I_C = 100$ µA, with a McCumber parameter $\beta_C = 1$ and $R_n I_C = 5.9$ mV. We can see that taking the quasiparticles contribution into account induces a delay of 1.1 ps in the generation of SFQ pulses.

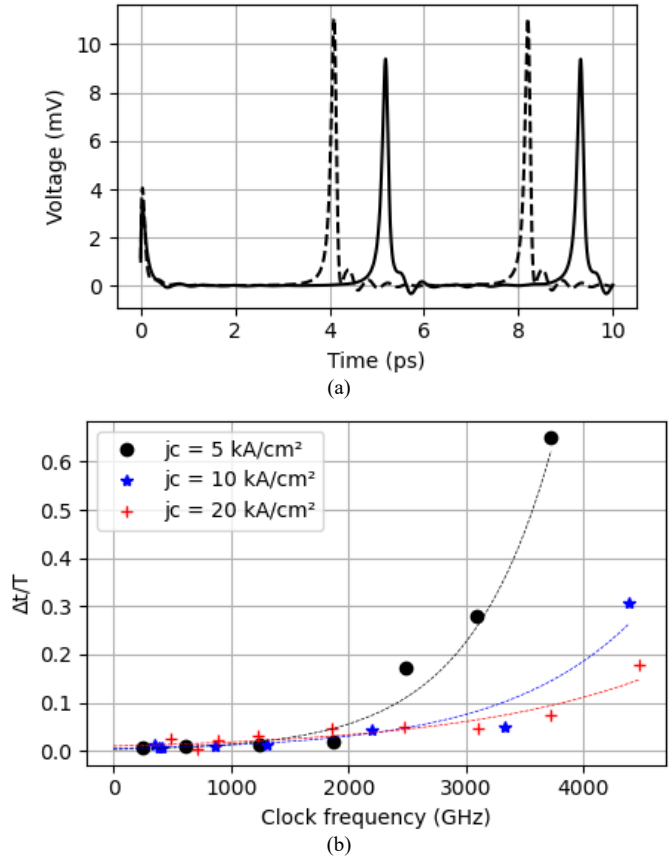

Fig. 5.  (a) Simulated generation of a SFQ voltage pulse train at 244 GHz for a current biased MgB$_2$-based Josephson junction for Cooper pairs alone (dashed line) or with quasiparticles contribution (solid line). The junction was current-biased just over $I_C = 100$ µA, with $j_C = 10\ kA/cm^2$, $\beta_C = 1$ and $R_n I_C = 5.9$ mV. (b) Simulated evolution of the normalized delay induced by the quasiparticles in the SFQ pulses generation as a function of the frequency of the generated SFQ pulse train, for 3 critical current densities. Markers are simulated points while lines are a guide for the eye. The induced delay is normalized to the SFQ signal period.

Fig. 5-b shows the evolution of this induced delay as a function of the clock frequency of the generated SFQ pulses, for three Josephson junctions with a critical current density of 5 kA/cm², 10 kA/cm² and 20 kA/cm² and an $R_n I_C$ product of 0.641 mV, 0.907 mV and 1.28 mV respectively, with the same $\beta_C = 1$, specific capacitance of 40 fF/µm² and area of 1 µm². One can see that the relative delay becomes higher when the SFQ signal frequency reaches the lower than the higher gap frequency due to dispersive effects. We may also see that this effect is more prominent at lower frequencies for lower critical current densities. This is consistent with the fact that pulse widths (of the order of 0.10 to 0.15 ps) are wider for lower current densities, corresponding to a lower spectral density at higher frequencies and hence to a lower ability to break Cooper pairs. The dispersion, associated with a higher pulse delay, occurs at about 3 THz for $j_C$ = 5 kA/cm² and at frequencies higher than 4 THz for higher current densities.

## V. Conclusion

The simulation tool developed in this work is helpful to predict the behavior of analog and digital devices and circuits with more accuracy, especially for high frequencies applications. It also enables simulations of non-conventional material-based Josephson junctions in the time-domain like MgB$_2$ junctions. It is shown that the behavior from simulations agrees well with well-known expectations, such as Shapiro steps and photo-assisted steps when a junction is irradiated by a microwave signal, or hysteresis in the *I-V* curve when the capacitance of the junction is considered.